\documentclass[journal]{IEEEtran}
\ifCLASSINFOpdf
\else
\fi

\usepackage{amsmath} 
\usepackage{amssymb}  
\usepackage{algorithmic}
\usepackage{mathtools}
\usepackage[ruled]{algorithm}
\usepackage{subfig}
\usepackage[usenames,dvipsnames]{xcolor}
\newcommand{\INDSTATE}[1][1]{\STATE\hspace{#1\algorithmicindent}}

 \newcommand{\sgn}{\mathop{\mathrm{sgn}}}
\newcommand{\sat}{\mathop{\mathrm{sat}}}

\definecolor{myred}{rgb}{0.75,0.0,0.1}

\begin{document}
%
\title{Reinforcement learning for port-Hamiltonian systems}
%
%
%


\author{Olivier~Sprangers,~
        Gabriel~A. D.~Lopes,~
        and~Robert~Babu\v{s}ka
\thanks{Sprangers, Lopes, and Babu\v{s}ka are with the Delft Center for Systems \& Control, Delft University of Technology, Mekelweg 2, 2628 CD Delft, The Netherlands. Email: osprangers@gmail.com; \{g.a.delgadolopes, r.babuska\}@tudelft.nl}
\thanks{Manuscript received XXXX, 2013; revised XXXX, 2013.}}

%
%

\markboth{XXXXXXXXX,~Vol.~XX, No.~XX, XXXXX~2013}%
{Sprangers \MakeLowercase{\textit{et al.}}: Reinforcement learning for port-Hamiltonian systems}

%



\maketitle

\begin{abstract}
Passivity-based control (PBC) for port-Hamiltonian systems provides an intuitive way of achieving stabilization by rendering a system passive with respect to a desired storage function. However, in most instances the control law is obtained without any performance considerations and it has to be calculated by solving a complex partial differential equation (PDE). In order to address these issues we introduce a reinforcement learning approach into the energy-balancing passivity-based control (EB-PBC) method, which is a form of PBC in which the closed-loop energy is equal to the difference between the stored and supplied energies. We propose a technique to parameterize EB-PBC that preserves the systems's PDE matching conditions, does not require the specification of a global desired Hamiltonian, includes performance criteria, and is robust to extra non-linearities such as control input saturation. The parameters of the control law are found using actor-critic reinforcement learning, enabling learning near-optimal control policies satisfying a desired closed-loop energy landscape. The advantages are that near-optimal controllers can be generated using standard energy shaping techniques and that the solutions learned can be interpreted in terms of energy shaping and damping injection, which makes it possible to numerically assess stability using passivity theory. From the reinforcement learning perspective, our proposal allows for the class of port-Hamiltonian systems to be incorporated in the actor-critic framework, speeding up the learning thanks to the resulting parameterization of the policy. The method has been successfully applied to the pendulum swing-up problem in simulations and real-life experiments. 
\end{abstract}


%
\IEEEpeerreviewmaketitle

\section{Introduction}
%
%
%
%
\IEEEPARstart{P}{assivity}-based control (PBC) \cite{Ortega1989} is a
methodology that achieves the control objective by rendering a system passive
with respect to a desired storage function \cite{Ortega2008}. Different forms
of PBC have been successfully applied to design robust controllers
\cite{Schaft2000} for mechanical systems and electrical circuits
\cite{Ortega2008,Ortega2001}. A key feature of PBC is that it exploits structural properties of the system. In
this paper, we are interested in the passivity-based control of systems endowed
with a special structure, called port-Hamiltonian (PH) systems. PH systems have
been widely used in PBC applications \cite{Duindam2009,Secchi2007}. Their
geometric structure allows reformulating a PBC problem in terms of solving a
set of partial differential equations (PDE's). Much research in the literature
concerns solving or simplifying such generally complex PDE's \cite{Ortega2008}. 

The drive for passivity-based control of port-Hamiltonian systems is grounded in the search for global stability, thus strongly relying on models. Other control techniques have been developed when no models are known and performance is important. One such example is reinforcement learning (RL) \cite{Sutton1998}. RL is a semi-supervised learning control method that
can solve optimal (stochastic) control problems for nonlinear systems, without
the need for a process model or for explicitly solving complex equations. In RL
the controller receives an immediate numerical reward as a function of the
process state and possibly control action. The goal is to find an optimal
control policy that maximizes the cumulative long-term rewards, which
corresponds to maximizing a value function \cite{Sutton1998}. In this paper, we
use actor-critic techniques \cite{Konda2003}, which are a class of RL methods
in which a separate actor and critic are learned. The critic approximates the
value function and the actor the policy (control law). Actor-critic
reinforcement learning is suitable for problems with continuous state and
action spaces. A general disadvantage of RL is that the progress of learning
can be very slow and non-monotonic. However, by incorporating (partial) model knowledge, learning can be sped up \cite{Grondman2011a}.

In this paper we address two important issues: First, we propose a learning control structure within the PH framework that retains important properties of the PBC. To this end, first a parameterization of a
particular type of PBC, called energy-balancing passivity-based control
(EB-PBC), is proposed such that the PDE arising in EB-PBC can be split into a
non-assignable part satisfying a matching condition following from the EB-PBC
framework and an assignable part that can be parameterized. Then, by applying
actor-critic reinforcement learning the parameterized part can be learned while automatically verifying the matching PDE. 
This can be seen as a paradigm shift from the traditional model-based control synthesis for PH systems: we do not seek to synthesize a controller in closed-form, but we aim instead to learn one online with proper structural constraints. This brings a number of advantages: I) It allows to specify the control goal in a ``local'' fashion through a reward function, without having to consider the entire global behavior of the system. The simplest example to illustrate this idea is by considering a reward function to be 1 when the system is in a small neighborhood of the desired goal and 0 everywhere else \cite{Sutton1998}. The learning algorithm will eventually find a global control policy. In the model-based PBC synthesis counterpart one needs to specify a desired global Hamiltonian. II) Learning brings performance in addition to the intrinsic stability properties of PBC. The structure of RL is such that the rewards are maximized, and these can include performance criteria, such as minimal time, energy consumption, etc. III) Learning offers additional robustness and adaptability since it tolerates model uncertainty in the PH framework.

From a learning control point of view, we present a systematic way of incorporating a priori knowledge into the RL problem. The approach proposed in this paper yields, after learning, a controller that can be interpreted in terms of energy shaping control strategies. The same interpretability is typically not found in the traditional RL solutions.


Thus, this work combines the advantages of both aforementioned control techniques, PBC and RL, and mitigates some of
their respective disadvantages. Historically, the trends in control synthesis have oscillated between performance and stabilization. PBC of PH systems is rooted in the stability of multi-domain nonlinear systems. By including learning we aim to address performance in the PH framework. In the experimental section of the paper, we show that our method is also robust to unmodeled nonlinearities, such as control input saturation. Control input saturation in PBC for PH systems has been addressed explicitly in the literature \cite{wei10,Astrom2008,Escobar1999,Fujimoto2003,Macchelli02,Macchelli03,sun09}. We show that our approach solves the problem of control input saturation on the learning side without the need of augmenting the model-based PBC. 

The work presented in this paper draws an interesting parallel with the application of iterative feedback tuning (IFT) \cite{Hjalmarsson02} in the PH framework \cite{Fujimoto08}. Both techniques optimize the parameters of the controller online, with the difference that in  IFT the objective is to minimize the error between the desired output and the measured output of the system, while our approach aims at maximizing a reward function, that can be very general. The choice of RL is warranted by its semi-supervised paradigm, as opposed to other traditional fully-supervised learning techniques, such as artificial neural networks or fuzzy approximators where the control specification (function approximation information) is input/output data instead of reward functions. Such fully-supervised techniques can be used within RL as function approximators to represent the value functions and control policies. Genetic algorithms can also be considered as an alternative to RL since they rely on fitness functions that are analogous to the rewards functions of RL. We aim to explore such classes of algorithms in our future work.

The theoretical background on PH systems and actor-critic reinforcement
learning is described in Section~\ref{sec:PH} and Section~\ref{sec:RL},
respectively. In Section~\ref{sec:PR}, our proposal for a parameterization of
input-saturated EB-PBC control, compatible with actor-critic reinforcement
learning, is introduced.  We then specialize this result to mechanical systems in Section~\ref{sec:MS}. Section~\ref{sec:RE} provides simulation and experimental results for the problem of swinging up an input-saturated inverted
pendulum 
and  Section~\ref{sec:CO} concludes the paper.

\section{Port-Hamiltonian Systems} \label{sec:PH}
Port-Hamiltonian (PH) systems are a natural way of representing a physical system in terms of its energy exchange with the environment through ports \cite{Ortega2001}. The general framework of PH systems was introduced in \cite{Maschke1992} and was formalized in \cite{Schaft1994,Schaft2000}. In this paper, we consider the input-state-output representation of the PH system which is of the form\footnote{We use the notation $\nabla_x \coloneqq \partial / \partial x$. Furthermore, all (gradient) vectors are column vectors.}:
\begin{equation}
\Sigma: \left\{
\begin{split}
\dot{x} &= \left[J(x)-R(x)\right]\nabla_x H(x) + g(x)u \\
y	   &= g^T(x)\nabla_x H(x) \end{split} \right. \label{eq:PHsystem}
\end{equation}
where $x \in \mathbb{R}^n$ is the state vector, $u \in \mathbb{R}^m,~m\le n$ is
the control input, $J,R: \mathbb{R}^n \to \mathbb{R}^{n \times n}$ with
$J(x) = -J(x)^T$ and $R(x) = R(x)^T \ge 0$ are the interconnection and damping
matrix, respectively, $H: \mathbb{R}^n \to \mathbb{R}$ the Hamiltonian
which is the stored energy in the system, $u,y \in \mathbb{R}^m$ are conjugated
variables whose product has the unit of power and $g: \mathbb{R}^n \to
\mathbb{R}^{n \times m}$ is the input matrix assumed to be full rank. For the
remainder of this paper, we denote:
\begin{equation}
F(x) \coloneqq J(x)-R(x)
\end{equation}
This matrix satisfies $F(x)+F^T(x) = -2R(x) \le 0$. System \eqref{eq:PHsystem} satisfies the power-balance equation \cite{Ortega2008}:
\begin{align}
\dot{H}(x) &=  \left(\nabla_x H(x)\right)^T \dot{x} \nonumber \\
		&=-\left(\nabla_x H(x)\right)^TR(x) \nabla_x H(x) + u^Ty \label{eq:pwbalance}
\end{align}
Since $R(x) \ge 0$, we obtain:
\begin{equation}
\dot{H}(x) \le u^Ty \label{eq:cyclopass}
\end{equation}
which is called the passivity inequality, if $H(x)$ is positive semi-definite, and cyclo-passivity inequality, if $H(x)$ is not positive semi-definite nor bounded from below \cite{Ortega2008}. Hence, systems satisfying \eqref{eq:cyclopass} are called (cyclo-)passive systems. The goal is to obtain the target closed-loop system:
\begin{equation}
\Sigma_{\mathrm{cl}}: \dot{x} = [J(x)-R_\mathrm{d}(x)]\nabla_x H_\mathrm{d}(x) \label{eq:tdyn}
\end{equation}
through \emph{energy shaping} using EB-PBC \cite{Ortega2008} and \emph{damping injection}, such
that $H_\mathrm{d}(x)$ is the desired closed-loop energy which has a minimum at
the desired equilibrium $x^*$ and satisfies:
\begin{equation}
\dot{H}_\mathrm{d}(x) = -\left(\nabla_x H_\mathrm{d}(x)\right)^T R_\mathrm{d}(x) \nabla_x H_\mathrm{d}(x) \label{eq:dotHD}
\end{equation}
which implies (cyclo-)passivity according to
\eqref{eq:pwbalance}-\eqref{eq:cyclopass} if the desired damping
$R_\mathrm{d}(x) \ge 0$. Hence, the control objective is achieved by rendering
the closed-loop system passive with respect to the desired storage function
$H_\mathrm{d}(x)$.

\subsection{Energy Shaping}
Define the added energy function:
\begin{equation}
 H_\mathrm{a}(x) \coloneqq H_\mathrm{d}(x)-H(x) \label{eq:ha}
\end{equation}
A state-feedback law $u_{\mathrm{es}}(x)$ is said to satisfy the energy-balancing property if it satisfies:
\begin{equation}
\dot{H}_\mathrm{a}(x) = -u_{\mathrm{es}}^T(x)y \label{eq:dotadded}
\end{equation}
If \eqref{eq:dotadded} holds, the desired energy $H_\mathrm{d}(x)$ is the
difference between the stored and supplied energy. Assuming $g(x) \in
\mathbb{R}^{n \times m},~m < n$, rank $\{g(x)\}=m$, the control law:
\begin{equation}
u_{\mathrm{es}}(x) = g^\dagger(x) F(x)\nabla_x H_\mathrm{a}(x) \label{eq:ues}
\end{equation}
with $g^\dagger(x) = (g^T(x)g(x))^{-1}g^T(x)$ solves the EB-PBC problem with $H_\mathrm{a}(x)$ a solution of the following set of PDE's \cite{Ortega2008}:
\begin{equation}
\begin{bmatrix} g^\perp(x) F^T(x) \\ g^T(x) \end{bmatrix}\nabla_x H_\mathrm{a}(x) = 0 \label{eq:ebpbcpde}
\end{equation}
with $g^\perp(x) \in \mathbb{R}^{(n-m) \times n}$ the full rank left-annihilator of $g(x)$, i.e. $g^\perp(x)g(x) = 0$.

\subsection{Damping Injection}
Damping is injected by feeding back the (new) passive output $g^T(x) \nabla_x H_\mathrm{d}(x)$,
\begin{equation}
u_{\mathrm{di}}(x) = -K(x)g^T(x) \nabla_x H_\mathrm{d}(x)
\end{equation}
with $K(x) \in \mathbb{R}^{m \times m}, K(x) = K^T(x) \ge 0$ such that:
\begin{equation}
R_\mathrm{d}(x) = R(x)+g(x)K(x)g^T(x) \label{eq:damping}
\end{equation}
Hence, the full control law consists of an energy shaping part and a damping injection part:
\begin{align}
u(x) &= u_{\mathrm{es}}(x) + u_{\mathrm{di}}(x) \nonumber \\
       &= g^\dagger(x) F(x)\nabla_x H_\mathrm{a}(x) \nonumber \\
		&\qquad - K(x)g^T(x) \nabla_x H_\mathrm{d}(x)   \label{eq:uebpbc}
\end{align}


\section{Actor-Critic Reinforcement Learning} \label{sec:RL}

In reinforcement learning, the system to be controlled (called 
`environment' in the RL literature) is modeled as a Markov decision process
(MDP). In a deterministic setting, this MDP is defined by the tuple
$M(X,U,f,\rho)$, where $X$ is the state space, $U$ the action space and $f: X
\times U \to X$ the state transition function that describes the process to
be controlled that returns the state $x_{k+1}$ after applying action $u_k$ in
state $x_k$. The vector $x_k$ is obtained by applying a zero-order hold
discretization $x_k = x(kT_s)$ with $T_s$ the sampling time.
The reward function is defined by $\rho: X \times U \to
\mathbb{R}$ and returns a scalar reward $r_{k+1} = \rho(x_{k+1},u_{k})$ after
each transition. The goal of RL is to find an optimal control policy $\pi : X
\to U$ by  maximizing an expected cumulative or total reward described as
some function of the immediate expected rewards. In this paper, we consider a
discounted sum of rewards. The value function $V^\pi: X \to \mathbb{R}$,
\begin{align}
V^\pi(x) &= \sum_{i=0}^\infty \gamma^i r^{\pi}_{k+i+1}\nonumber\\
&= \sum_{i=0}^\infty  \gamma^i\rho(x_{k+i+1},\pi(x_{k+i})), \quad x = x_k
\end{align}
approximates this discounted sum during learning while following policy $\pi$ where $\gamma \in [0,1)$ is the discount factor.

When dealing with large and/or continuous state and action spaces, it is
necessary to approximate the value function and policy. Actor-critic (AC)
algorithms \cite{Barto1983,Konda2003} learn a separate actor (policy $\pi$)
and critic (value function $V^\pi$). The critic approximates and updates
(improves) the value function. Then, the actor's parameters are updated in the
direction of that improvement. The actor and critic are usually defined by a
differentiable parameterization such that gradient ascent can be used to update
the parameters. This is beneficial when dealing with continuous action spaces
\cite{Sutton2000}.
In this paper, the temporal-difference based Standard Actor-Critic (S-AC)
algorithm from \cite{Grondman2011a} is used. Define the approximated policy
$\hat{\pi}: \mathbb{R}^n \to \mathbb{R}^m$ and the approximated value
function as $\hat{V}: \mathbb{R}^n \to \mathbb{R}$. Denote the
parameterization of the actor by $\vartheta \in \mathbb{R}^p$ and of the critic
by $\theta \in \mathbb{R}^q$. The temporal difference \cite{Sutton1998}:
\begin{equation}
\delta_{k+1} := r_{k+1} + \gamma \hat{V}(x_{k+1},\theta_k) - \hat{V}(x_{k},\theta_k) \label{eq:ebactempd}
\end{equation}
is used to update the critic parameters using the following gradient ascent update rule:
\begin{align}
\theta_{k+1} = \theta_k + \alpha_\mathrm{c} \delta_{k+1} \nabla_\theta \hat{V}(x_k,\theta_k)
\end{align}
in which $\alpha_\mathrm{c} > 0$ is the learning rate. Eligibility traces $e_k \in \mathbb{R}^q$ \cite{Sutton1998} can be used to speed up learning by including reward information about previously visited states. The update for the critic parameters becomes:
\begin{align}
e_{k+1} &= \gamma \lambda e_k +  \nabla_\theta \hat{V}(x_k,\theta_k) \\
\theta_{k+1} &= \theta_k + \alpha_c \delta_{k+1} e_{k+1}
\end{align}
with $\lambda \in [0,1)$ the trace-decay rate. The policy approximation can be
updated in a similar fashion, as described below.
RL needs exploration in order to visit new, unseen parts of the state-action
space so as to possibly find better policies. This is achieved by perturbing
the policy with a exploration term $\Delta u_k$. Many techniques have been developed for choosing the type of exploration term (see e.g. \cite{asmuth2009}). In this paper we consider $\Delta u_k$ to be random with zero-mean. In the experimental section we choose $\Delta u_k$ to be have a normal distribution. The overall control action now becomes:
\begin{equation}
u_k = \hat{\pi}(x_k,\vartheta_k) + \Delta u_k \label{eq:overallaction}
\end{equation}
The policy update is such that the policy parameters are updated towards
(away from) $\Delta u_k$ if the temporal difference \eqref{eq:ebactempd} is positive (negative).
This leads to the following policy update rule:
\begin{equation}
\vartheta_{k+1} = \vartheta_k + \alpha_\mathrm{a} \delta_{k+1} \Delta u_k \nabla_\vartheta \hat{\pi}(x_k,\vartheta_k) \label{eq:actorpa}
\end{equation}
with $\alpha_\mathrm{a} > 0$ the actor learning rate.

\section{Energy-Balancing Actor-Critic} \label{sec:PR} In this section we
present our main results. Our approach is that we will use the PDE
\eqref{eq:ebpbcpde} and split it into an assignable, parameterizable part and
an unassignable part that satisfies the matching condition. In this way, it is
possible to parameterize the desired closed-loop Hamiltonian $H_\mathrm{d}(x)$
and simultaneously satisfy \eqref{eq:ebpbcpde}. After that, we parameterize the
damping matrix $K(x)$. The two parameterized variables --- the desired
closed-loop energy $H_\mathrm{d}(x)$ and damping $K(x)$ --- are then suitable for
Actor-Critic RL by defining two actors for these variables.
First, we reformulate the PDE \eqref{eq:ebpbcpde} in terms of the desired closed-loop energy $H_\mathrm{d}(x)$ by applying \eqref{eq:ha}:
\begin{align}
\underbrace{\begin{bmatrix} g^\perp(x) F^T(x) \\ g^T(x) \end{bmatrix}}_{A(x)}\left(\nabla_x H_\mathrm{d}(x) - \nabla_x H(x)\right) = 0 \label{eq:pdeebpbc2}
\end{align}
and we denote the kernel of $A(x)$ as:
\begin{equation}
\ker(A(x)) = \{ N(x) \in \mathbb{R}^{n \times b} : A(x) N(x) = 0 \}
\end{equation}
such that \eqref{eq:pdeebpbc2} reduces to:
\begin{equation}
\nabla_x H_\mathrm{d}(x) - \nabla_x H(x) = N(x)a
\end{equation}
with $a \in \mathbb{R}^b$. Suppose that (an example is given further on) the state vector $x$ can be split, such that $x = [w^T~  z^T]^T$, where $z \in \mathbb{R}^{c}$ and $w \in \mathbb{R}^d,~c+d=n$ corresponding to the zero and non-zero elements of $N(x)$ such that:
\begin{equation}
\begin{bmatrix} \nabla_{w} H_\mathrm{d}(x) \\ \nabla_{z} H_\mathrm{d}(x) \end{bmatrix} - \begin{bmatrix} \nabla_{w} H(x) \\ \nabla_{z} H(x) \end{bmatrix} = \begin{bmatrix} N_w(x) \\ 0 \end{bmatrix} a \label{eq:pdesplit}
\end{equation}
We assume that the matrix $N_{w}(x)$ is rank $d$, which is always true for fully actuated mechanical systems (see Section \ref{sec:MS}). It is clear that $\nabla_{z} H_\mathrm{d}(x) = \nabla_{z} H(x)$, which we call
the matching condition, and hence $\nabla_{z} H_\mathrm{d}(x)$ cannot be chosen
freely. Thus, only the desired closed-loop energy gradient vector $\nabla_{w}
H_\mathrm{d}(x)$ is free for assignment. We consider a $\xi$-parameterized total desired energy with the following form:
\begin{align}
\hat{H}_\mathrm{d}(x,\xi)  &:=H(x) + \xi^T \phi_H(w) +\bar{H}_\mathrm{d}(w)+ C \label{eq:deshampar}
\end{align}
where $ \xi^T \phi_H(w) $ represents a linear-in-parameters basis function approximator ($\xi \in \mathbb{R}^e$ a parameter vector and $\phi_H(w) \in \mathbb{R}^e$ the basis function, with $e$ chosen sufficiently large to represent the assignable desired closed-loop energy), $\bar{H}_\mathrm{d}(w)$ is an arbitrary function of $w$,
 and $C$ chosen to render $\hat{H}_\mathrm{d}(x,\xi)$ non-negative. The function $\hat{H}_\mathrm{d}(x,\xi)$ automatically verifies (\ref{eq:pdesplit}). To guarantee that the system is passive in relation to the storage function $\hat{H}_\mathrm{d}(x,\xi)$ the basis functions $\phi_H(w)$ should be chosen to be bounded such that  $\xi^T\phi_H(w)$ does not grow unbounded towards $-\infty$ when $||w||\to \infty$. Moreover, it is important to constrain the minima of $\hat{H}_\mathrm{d}(x,\xi)$ to be the desired equilibrium $x^*$, via the choice of the basis functions.
%
%
%
%
%
%
%
The elements of the desired damping matrix $K(x)$ of \eqref{eq:uebpbc}, denoted $\hat{K}(x,\Psi)$, can be parameterized in a similar way:
\begin{equation}
[\hat{K}(x,\Psi)]_{ij} = \sum_{l=1}^f [\Psi]_{ijl} [\phi_K(x)]_l \label{eq:gab1}
\end{equation}
with $\Psi \in \mathbb{R}^{m \times m \times f}$ and
\begin{equation}
[\Psi]_{ijl} = [\Psi]_{jil} \label{eq:psisym}
\end{equation}
a parameter vector such that $\hat{K}(x,\Psi) = \hat{K}^T(x,\Psi)$, $(i,j)=1,\dots,m$ and $\phi_K(x) \in \mathbb{R}^f$ basis functions. We purposefully do not impose $\hat{K}(x,\Psi) \ge 0$ to allow the injection of energy in the system via the damping term. This idea has been used in \cite{Koopman:2012tm}  to overcame the dissipation obstacle when synthesizing controllers by interconnection. Although this breaches the passivity criterion of \eqref{eq:cyclopass} in our particular case we show that local stability can still be numerically demonstrated using passivity analysis in Section~\ref{subsec:stab}. This choice is made based on the knowledge that the standard Energy Balancing PBC method (without any extra machinery to accommodate saturation) cannot stabilize in the up position a saturated-input pendulum system starting from the down position. As such, this choice illustrates the power of RL in finding alternative routes to obtain control policies, such as injecting energy though the damping term. In other settings enforcing that $\hat{K}(x,\Psi) \ge 0$ benefits the stability analysis. The control law \eqref{eq:uebpbc} now becomes (when no ambiguity is present, the function arguments are dropped to improve readability):
\begin{align}
u(x,\xi,\Psi)& = g^\dagger(x) F(x)\begin{bmatrix}\nabla_{w} \hat{H}_\mathrm{d}(x,\xi) - \nabla_{w} H(x) \\ 0  \end{bmatrix} \nonumber \\
				& \quad - \hat{K}(x,\Psi)g^T(x) \nabla_x \hat{H}_\mathrm{d}(x,\xi) \nonumber \\
		&= g^\dagger F\begin{bmatrix}  D_w^T \phi_H \xi
		+ \nabla_w \bar{H}_{\mathrm{d}}  
		\\ 0  \end{bmatrix} \nonumber \\
				& \quad - \hat{K}g^T \begin{bmatrix} D_w^T \phi_H \xi 
				+ \nabla_w \bar{H}_{\mathrm{d}}+ \nabla_w{H}
				 \\ \nabla_{z} H  \end{bmatrix} \label{eq:uparam}
\end{align}
Now, we are ready to introduce the update equations for the parameter vectors $\xi,~[\Psi]_{ij}$. Denote by $\xi_k,~[\Psi_k]_{ij}$ the value of the parameters at the discrete time step $k$. The policy $\hat{\pi}$ of the actor-critic reinforcement learning algorithm is chosen equal to the control law parameterized by \eqref{eq:uparam}:
\begin{equation}
\hat{\pi}(x_k,\xi_k,\Psi_k) \coloneqq u(x_k,\xi_k,\Psi_k)
\end{equation}
In this paper we take the control input saturation problem into account by considering a generic saturation function $\varsigma: \mathbb{R}^m \to S,~S\subset \mathbb{R}^m$, such that:
\begin{equation}
 \varsigma(u(x)) \in S ~\forall u \label{eq:saturation}
\end{equation}
where $S$ is the set of valid control inputs. The control action with exploration
\eqref{eq:overallaction} becomes:
\begin{equation}
u_k = \varsigma\left(\hat{\pi}(x_k,\xi_k,\psi_k) + \Delta u_k\right) \label{eq:overallaction2}
\end{equation}
where $\Delta u_k$ is drawn from a desired distribution. The exploration term to be used in the actor update \eqref{eq:actorpa} must be adjusted to respect the saturation:
\begin{equation}
\Delta \bar{u}_k = u_k - \hat{\pi}(x_k,\xi_k,\psi_k)
\end{equation}
Note that due to this step, the exploration term $\Delta\bar{u}_k$ used in the learning algorithm is no longer drawn from the chosen distribution present in $\Delta u_k$. Furthermore, we obtain the following gradients of the saturated policy:
\begin{align}
\nabla_\xi \varsigma(\hat{\pi}) &= \nabla_{\hat{\pi}} \varsigma(\hat{\pi}) \nabla_\xi \hat{\pi} \label{eq:polgradsat} \\
\nabla_{[\Psi]_{ij}} \varsigma(\hat{\pi}) &= \nabla_{\hat{\pi}} \varsigma(\hat{\pi}) \nabla_{[\Psi]_{ij}} \hat{\pi}  \label{eq:polgradsat1}
\end{align}
Although not explicitly indicated in the previous equations, the (lack of) differentiability of the saturation function $\varsigma$ has to be considered for the problem at hand such that the computation of the gradient of $\varsigma$ can be made. For a traditional saturation in $u_i\in \mathbb{R}$ of the form $\max(u_{min},\min(u_{max},u_i))$, i.e. assuming each input $u_i$ is bounded by $u_{min}$ and $u_{max}$, then the gradient of $\varsigma$ is the zero matrix outside the unsaturated set $S$ (i.e. when $u_i < u_{min}$ or $u_i > u_{max}$). For other types of saturation the function $\nabla \varsigma$ must be computed. Finally, the actor parameters $\xi_k, [\Psi_k]_{ij}$ are updated according to \eqref{eq:actorpa}, respecting the saturated policy gradients. For the parameters of the desired Hamiltonian we obtain:
\begin{equation}
\xi_{k+1} = \xi_{k} + \alpha_{\mathrm{a},\xi} \delta_{k+1} \Delta \bar{u}_k  \nabla_{\xi} \varsigma \left(\hat{\pi}(x_k,\xi_k,\Psi_k)\right) \label{eq:actorhd}
\end{equation}
and for the parameters of the desired damping we have:
\begin{align}
 [\Psi_{k+1}]_{ij} &= [\Psi_k]_{ij}+  \nonumber \\
			&\alpha_{\mathrm{a},[\Psi]_{ij}} \delta_{k+1} \Delta \bar{u}_k \nabla_{[\Psi_k]_{ij}} \varsigma \left(\hat{\pi}(x_k,\xi_k,\Psi_k)\right) \label{eq:actorrd}
\end{align}
where $(i,j)=1,\dots,m$, while observing \eqref{eq:psisym}.
Algorithm~\ref{alg:1} gives the entire Energy-Balancing Actor-Critic
algorithm with input saturation.
\begin{algorithm*}[t!]
\caption{Energy-Balancing Actor-Critic}
\label{alg:1}
\begin{algorithmic}[1]
\REQUIRE System \eqref{eq:PHsystem}, $\lambda$, $\gamma$, $\alpha_\mathrm{a}$ for each actor, $\alpha_\mathrm{c}$.
\STATE $e_0(x) = 0 \qquad \forall x$
\STATE Initialize $x_0$, $\theta_0$, $\xi_0$, $\Psi_0$
\STATE $k \gets 1$
\STATE \textbf{loop}
\INDSTATE[2] \textbf{Execute:}
\INDSTATE[3] Draw $\Delta u_k \sim \mathcal{N}(0,\sigma^2)$, calculate action $u_k = \varsigma\left(\hat{\pi}(x_k,\xi_k,\psi_k) + \Delta u_k\right)$, $\Delta \bar{u}_k =  u_k - \hat{\pi}(x_k,\xi_k,\psi_k)$
\INDSTATE[3] Observe next state $x_{k+1}$ and calculate reward $r_{k+1} = \rho(x_{k+1},u_k)$
	\INDSTATE[2] \textbf{Critic:}
	\INDSTATE[3] Temporal difference: $\delta_{k+1} = r_{k+1} + \gamma \hat{V}(x_{k+1},\theta_k) - \hat{V}(x_{k},\theta_k)$
	\INDSTATE[3] Eligibility trace: $e_{k+1} =  \gamma \lambda e_k +  \nabla_\theta \hat{V}(x_k,\theta_k) $
	\INDSTATE[3] Critic update: $\theta_{k+1} = \theta_k + \alpha_\mathrm{c} \delta_{k+1} e_{k+1}$
	\INDSTATE[2] \textbf{Actors:}
	\INDSTATE[3] Actor 1 ($\hat{H}_\mathrm{d}(x,\xi)$): $\xi_{k+1}=\xi_{k} + \alpha_{\mathrm{a},\xi} \delta_{k+1} \Delta \bar{u}_k  \nabla_{\xi} \varsigma \left(\hat{\pi}(x_k,\xi_k,\Psi_k)\right)$
	\INDSTATE[3] Actor 2 ($\hat{K}(x,\Psi)$):
	\INDSTATE[4] \textbf{for $i,j=1,\dots,m$ do}
	\INDSTATE[5] $[\Psi_{k+1}]_{ij} = [\Psi_k]_{ij}+ \alpha_{\mathrm{a},[\Psi]_{ij}} \delta_{k+1} \Delta \bar{u}_k \nabla_{[\Psi_k]_{ij}} \varsigma \left(\hat{\pi}(x_k,\xi_k,\Psi_k)\right)$
	\INDSTATE[4] \textbf{end for}
\STATE \textbf{end loop}
\end{algorithmic}
\end{algorithm*}

The dynamics of the Energy-Balancing Actor Critic Algorithm~\ref{alg:1} raises a number of questions regarding stability and convergence: are the good stability properties of the traditional energy-balancing PBC lost? In effect this is not the case, as if the parameter $\xi$ is fixed then stability is preserved, in the sense that the system is passive to the storage function $\hat{H}_\mathrm{d}(x,\xi)$, assuming bounded basis functions describing $\hat{H}_\mathrm{d}(x,\xi)$ as discussed after equation (\ref{eq:deshampar}) and that the dissipation matrix $\hat{K}$ is semi-positive definite. A related question is  if during learning (while the parameter $\xi$ is evolving) will the Hamiltonian $\hat{H}_\mathrm{d}(x,\xi)$ capture the desired control specification. One cannot assume that the desired Hamiltonian will immediately fulfil the control specification, since if that was the case then no learning is needed. In the RL community it is generally accepted that during learning no stability and convergence guarantees can be given \cite{Sutton1998}, as exploration is a necessary component of the framework. In our framework, we cannot guaranteed convergence during learning, but by constraining the desired Hamiltonian we can prevent the total energy to grow unbounded, avoiding possible instabilities. Another relevant question is: will RL converge in this setting?  The control law in (\ref{eq:uparam}) can be rewritten as 
\begin{equation*}
u=\bar{\phi}_1(x)+\sum_{ij}\xi_i \bar{\phi}_{2,i}(x)+
\bar{\Psi}_j \bar{\phi}_{3,j}(x)+\xi_i \bar{\Psi}_j\bar{\phi}_{4,ij}(x)
%
\end{equation*}
with $\bar{\Psi}$ representing the stacked version of $\Psi$, meaning that the policy is parameterised in an affine bilinear way, as opposed to the standard linear in parameter representations found in the standard actor-critic literature. As such, we cannot at this moment take advantage of the existing  actor-critic RL convergence proofs since we violate this condition. Given this hurdle we observe, however, that in practice not only will the RL component converge very fast (faster then traditional model-free RL) but throughout learning the system never gets unstable, as presented next in Section \ref{sec:RE}. Additionally, the resulting policy performs as well as standard model-free approximated RL algorithms.  The last point to consider is that since the Energy Balancing PBC suffers from the dissipation obstacle \cite{Ortega2008}, limiting its applicability to special classes of systems such as mechanical systems, the algorithm we present contains the same limitation. Eliminating such limitation is ongoing work (see also the results presented in \cite{Koopman:2012tm})

\section{Mechanical Systems}\label{sec:MS}
To illustrate an application of the method, consider a fully actuated mechanical system of the form:
\begin{equation}
\Sigma_{\mathrm{m}}: \left\{
\begin{split}
\begin{bmatrix} \dot{q} \\ \dot{p} \end{bmatrix} &= \begin{bmatrix} 0 &I \\ -I &-\bar{R} \end{bmatrix}\begin{bmatrix}\nabla_q H(q,p) \\ \nabla_p H(q,p) \end{bmatrix} + \begin{bmatrix} 0 \\ I \end{bmatrix} u \\
y &= \begin{bmatrix} 0 &I \end{bmatrix}\begin{bmatrix}\nabla_q H(q,p) \\ \nabla_p H(q,p) \end{bmatrix}
\end{split}
\right.
\label{eq:phmech}
\end{equation}
with $q \in \mathbb{R}^{\bar{n}},~ p \in \mathbb{R}^{\bar{n}}$ ($\bar{n} =
\frac{n}{2}$, $n$ even) the generalized positions and momenta, respectively, 
and $\bar{R} \in \mathbb{R}^{\bar{n} \times \bar{n}}$ the damping matrix. The system admits \eqref{eq:PHsystem} with $\bar{R} > 0$ and the
Hamiltonian:
\begin{equation}
H(q,p) = \frac{1}{2}p^T M^{-1}(q)p + P(q) \label{eq:hamiltonianmech}
\end{equation}
with $M(q) = M^T(q) >0$ the inertia matrix and $P(q)$ the potential energy. For
the system \eqref{eq:phmech} it holds that rank $\{g(x)\}=\bar{n}$ and the
state vector can be split into part $w = [q_1, q_2,\dots,q_{\bar{n}}]^T$ and part
$z = [p_1,p_2,\dots,p_{\bar{n}}]^T$. Since $g(x)=[0~I]^{T}$ its annihilator can be written as $g^{\perp}(x)=[\bar{g}(x)~0]$, for an arbitrary matrix $\bar{g}(x)$. 
This means that only the potential energy can be shaped, which is widely known in EB-PBC for mechanical systems. The approximated desired closed-loop energy \eqref{eq:deshampar} reads:
\begin{align}
\hat{H}_\mathrm{d}(x,\xi)  &=  \frac{1}{2}p^T M^{-1}(q)p + \xi^T \phi_H(q) \label{eq:deshamparmech}
\end{align}
where the first term represents the unassignable part, i.e. the kinetic energy of the system Hamiltonian \eqref{eq:hamiltonianmech}, and the second term $\xi^T \phi_H(q)$ the assignable desired potential energy.
%
%
The actor updates can be defined for each parameter according to
\eqref{eq:actorhd}--\eqref{eq:actorrd}. For underactuated mechanical systems,
e.g. $G = \left[0~I\right]^T$, the split state vector $z$ is enlarged with those $q$-coordinates that cannot be actuated directly because these coordinates correspond to the zero elements of $N(x)$, (i.e. the matrix $N_q(x)$ is no longer rank $\bar{n}$).


\section{Example: Pendulum Swing-up}\label{sec:RE}

\begin{figure}[htbp]
      \centering
      \includegraphics[width=0.85\columnwidth]{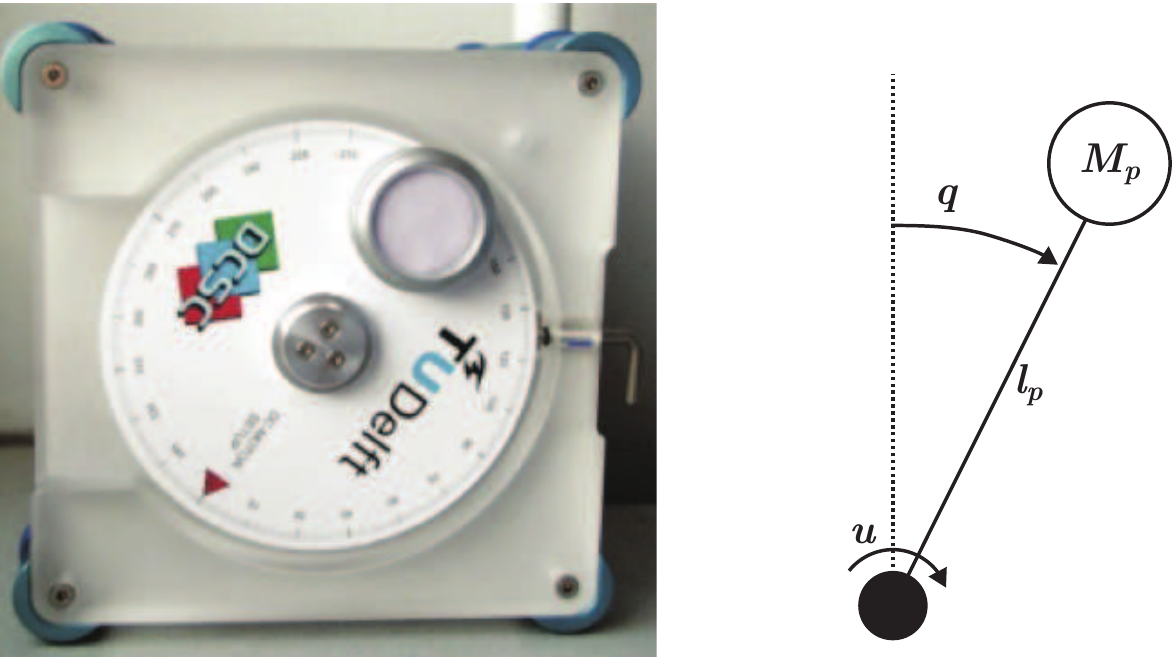}
      \caption[Inverted pendulum setup]{Inverted pendulum setup.}
      \label{fig:pendulum}
   \end{figure}

To validate our method, the problem of swinging up an inverted pendulum subject
to control saturation is studied in simulation and using the actual physical setup
depicted in Fig.~\ref{fig:pendulum}.

The pendulum swing-up is a low-dimensional, but highly nonlinear control
problem commonly used as a benchmark in the RL literature \cite{Grondman2011a}
and it has also been studied in PBC \cite{Astrom2008}. The equations of motion
admit \eqref{eq:phmech} and read:
\begin{equation}
\Sigma_{\mathrm{p}}:
\left\{
\begin{split}
\begin{bmatrix} \dot{q} \\ \dot{p} \end{bmatrix} &= \begin{bmatrix} 0 &1 \\ -1 &-\bar{R}(\dot{q}) \end{bmatrix}\begin{bmatrix}\nabla_q H(q,p) \\ \nabla_p H(q,p) \end{bmatrix} + \begin{bmatrix} 0 \\ \frac{K_\mathrm{p}}{R_\mathrm{p}} \end{bmatrix} u \\
y &=\begin{bmatrix} 0 &\frac{K_\mathrm{p}}{R_\mathrm{p}} \end{bmatrix}\begin{bmatrix}\nabla_q H(q,p) \\ \nabla_p H(q,p) \end{bmatrix}
\end{split}
\right.
\label{eq:pendsystem}
\end{equation}
with $q$ the angle of the pendulum and $p$ the angular momentum, thus we denote the full measurable state $x = [q,p]^T$.
The damping term is:
\begin{equation}
\bar{R}(\dot{q}) = b_\mathrm{p} + \frac{K_\mathrm{p}^2}{R_\mathrm{p}}+\frac{\sigma_\mathrm{p}}{\left\vert \dot{q} \right\vert}
\end{equation}
for which it holds that $\bar{R}(\dot{q}) > 0,\; \forall \dot{q}$. Note that the fraction $\sigma_\mathrm{p}/|\dot{q}|$ arrises from the modelled Coulomb friction $f_c=\sigma_\mathrm{p} ~\mathrm{sign}(\dot{q})=\sigma_\mathrm{p}/|\dot{q}|  \dot{q}=\sigma_\mathrm{p}/|\dot{q}|~ p/J_\mathrm{p}$.  
Furthermore, we denote the Hamiltonian:
\begin{equation}
H(q,p) = \frac{p^2}{2J_\mathrm{p}} + P(q) \label{eq:hamiltonianpend}
\end{equation}
with:
\begin{equation}
P(q) = M_\mathrm{p}g_\mathrm{p}l_\mathrm{p} (1+\cos q)
\end{equation}
The model parameters are given in Table~\ref{tab:modparam}.
\begin{table}[htbp]
\caption{Inverted pendulum model parameters}
\label{tab:modparam}
\begin{center}
\begin{tabular}{lcll}
\hline
\textbf{Model parameters} &\textbf{Symbol} &\textbf{Value} &\textbf{Units} \\
\hline
Pendulum inertia	&$J_\mathrm{p}$ &$1.90 \cdot 10^{-4}$ &kgm$^2$ \\
Pendulum mass	&$M_\mathrm{p}$		 &$5.2 \cdot 10^{-2}$ &kg \\
Gravity		&$g_\mathrm{p} $ &$9.81$ &m/s$^2$ \\
Pendulum length	&$l_\mathrm{p}$		 &$4.20 \cdot 10^{-2}$ &m \\
Viscous friction      &$b_\mathrm{p}$		 &$2.48 \cdot 10^{-6}$ &Nms \\
Coulomb friction           &$\sigma_\mathrm{p}$ 	 	&$1.0 \cdot 10^{-3}$ &N \\
Torque constant	&$K_\mathrm{p}$ 		&$5.60 \cdot 10^{-2}$ &Nm/A \\
Rotor resistance	&$R_\mathrm{p}$ &$9.92$		 &$\Omega$ \\
\hline
\end{tabular}
\end{center}
\end{table}
%
The desired Hamiltonian \eqref{eq:deshampar} reads:
\begin{align}
\hat{H}_\mathrm{d}(x,\xi) &=  \frac{p^2}{2J_\mathrm{p}}+ \xi^T \phi_H(q)   \label{eq:deshampar2}
\end{align}
Only the potential energy can be shaped that we denote by $\hat{P}_\mathrm{d}(q,\xi)= \xi^T \phi_H(q)$.
%
Furthermore, as there is only one input, $\hat{K}(x,\Psi)$ becomes a scalar:
\begin{equation}
\hat{K}(x,\psi) = \psi^T\phi_K(x)
\end{equation}
Thus, control law \eqref{eq:uparam} results in:
\begin{align}
u(x,\xi,\psi) &= g^\dagger F
\begin{bmatrix} 
\xi^T\nabla_q\phi_H-\nabla_q P\\0
\end{bmatrix}-\hat{K}g^T
\begin{bmatrix} 
\xi^T\nabla_q\phi_H
\\J_\mathrm{p}^{-1} p
\end{bmatrix}\nonumber\\
&= -\frac{R_\mathrm{p}}{K_\mathrm{p}}\left(\xi^T\nabla_q\phi_H +M_\mathrm{p}g_\mathrm{p}l_\mathrm{p}\sin(q)\right)\\
&~~~~~~~~ - \frac{K_\mathrm{p}}{R_\mathrm{p}} \psi^T\phi_K \dot{q}\label{eq:uparam2-2}
\end{align}
%
%
which we define as the policy $\hat{\pi}(x,\xi,\psi)$.
Hence, we have two actor updates:
\begin{align}
\xi_{k+1} &=  \xi_{k} + \alpha_{\mathrm{a},\xi} \delta_{k+1} \Delta \bar{u}_k  \nabla_{\xi} \varsigma \left(\hat{\pi}(x_k,\xi_k,\psi_k)\right) \label{eq:polgrades} \\
\psi_{k+1} &= \psi_{k} + \alpha_{\mathrm{a},\psi} \delta_{k+1} \Delta \bar{u}_k \nabla_{\psi} \varsigma \left(\hat{\pi}(x_k,\xi_k,\psi_k)\right) \label{eq:polgraddi}
\end{align}
for the desired potential energy $\hat{P}_\mathrm{d}(q,\xi)$ and the desired damping $\hat{K}(x,\psi)$, respectively.

\subsection{Function Approximation} \label{sec:funapprox} To approximate the
critic and the two actors, function approximators are necessary. In this paper
we use the Fourier basis \cite{Konidaris2008} because of its ease of use, the
possibility to incorporate information about the symmetry in the system and the ability to ascertain properties useful for stability analysis of this specific problem. The periodicity of the function approximators obtained via a Fourier basis is compatible with the topology of the configuration space of the pendulum, defined to be $S^1\times\mathbb{R}$.
We define a multivariate
$N$th-order\footnote{`Order' refers to the order of approximation; `dimensions'
to the number of states in the system.} Fourier basis for $n$ dimensions as:
\begin{equation}
\phi_i(\bar{x}) =  \cos(\pi c_i^T \bar{x}),~~ i\in \{1,\dots,(N+1)^n\}
\end{equation}
with $c_i \in \mathbb{Z}^n$, which means that all possible $N+1$ integer values, or frequencies, are combined in a vector in $\mathbb{Z}^n$ to create a matrix $c \in \mathbb{Z}^{n \times (N+1)^n}$ containing all possible frequency combinations. For example,
\begin{equation}
c_1 = [0~0]^T, \;\; c_2 = [1~0]^T,~\ldots~,~c_{(3+1)^2} = [4~4]^T
\end{equation}
for a 3$^{\mathrm{rd}}$-order Fourier basis in 2 dimensions.
The state $x$ is scaled according to:
\begin{equation}
\bar{x}_i = \frac{x_i-x_{i,\min}}{x_{i,\max}-x_{i,\min}}\left(\bar{x}_{i,\max}-\bar{x}_{i,\min}\right) + \bar{x}_{i,\min} \label{eq:statescaling}
\end{equation}
for $i=1,\dots,n$ with $(\bar{x}_{i,\min}, \bar{x}_{i,\max}) = (-1,1)$.
Projecting the state variables onto this symmetrical range has several
advantages. First, this means that the policy will be periodic with period
$T=2$, such that it wraps around (i.e., modulo $2\pi$) and prevents
discontinuities at the boundary values of the angle ($x = [\pi \pm \epsilon
,p]$, $\epsilon$ very small). This way we are taking into consideration the topology of the system: for the inverted pendulum we have that $q\in \mathbb{S}^1$ and $p\in \mathbb{R}$.
 Second, learning will be faster because updating
the value function and policy for some $x$ also applies to the sign-opposite
value of $x$. Third, $\dot{\hat{P}}_\mathrm{d}(0,\xi) =0$ by the choice of parameterization, which
is beneficial for stability analysis. Although the momentum will now also be
periodic in the value function and policy, this is not a problem because the
value function and policy approximation are restricted to a domain and the
momentum itself is also restricted to the same domain using saturation.
We adopt the adjusted learning rate from \cite{Konidaris2008} such that:
\begin{equation}
\alpha_{\mathrm{a}_i,\xi} = \frac{\alpha_{\mathrm{a}_\mathrm{b},\xi}}{\Vert c_i \Vert_2}, \qquad \alpha_{\mathrm{a}_i,\psi} = \frac{\alpha_{\mathrm{a}_\mathrm{b},\psi}}{\Vert c_i \Vert_2} \label{eq:lrates}
\end{equation}
for $i=1,\dots,(N+1)^n$ with $\alpha_{\mathrm{a}_\mathrm{b},\xi}$,
$\alpha_{\mathrm{a}_\mathrm{b},\psi}$ the base learning rate for the two actors
(Table~\ref{tab:simparam}) and $\alpha_{\mathrm{a}_1,\xi} =
\alpha_{\mathrm{a}_\mathrm{b},\xi}$, $\alpha_{\mathrm{a}_1,\psi} =
\alpha_{\mathrm{a}_\mathrm{b},\psi}$ to avoid division by zero for $c_1 =
[0~0]^T$. Equation~\eqref{eq:lrates} implies that parameters corresponding to
basis functions with higher (lower) frequencies are learned slower (faster).
The parameterizations described above result in $\dot{\hat{H}}_\mathrm{d}(x^*,\xi)=0$ for all $\xi$, where $x^*=[0,0]^T$ is the goal state. This entails that the goal state is a critical point of the Hamiltonian throughout the learning process, in effect speeding up the RL convergence. 
  
\subsection{Simulation}\label{sec:simulationpaper}

The task is to learn to swing up and stabilize the pendulum from the initial
position pointing down  $x_0 =[\pi, 0]^T$ to the desired equilibrium position at
the top $x^*=[0,0]^T$. Since the control action is saturated, the system is not
able to swing up the pendulum directly, but rather it must swing back and
forth to build up momentum to eventually reach the equilibrium.
The reward function $\rho$ is defined such that it has its maximum in the desired unstable equilibrium and penalizes other states via:
\begin{equation}
\rho(x,u) = Q_\mathrm{r}\left(\cos (q)-1\right) - R_\mathrm{r}p^2
\end{equation}
with:
\begin{equation}
Q_\mathrm{r} = 25~,~R_\mathrm{r}=\frac{0.1}{J_p^2}
\end{equation}
This reward function is consistent with the mapping $S^1 \to \mathbb{R}$ for the angle and proved to improve performance over a purely quadratic reward, such as the one used in e.g.~\cite{Grondman2011a}.
For the critic, we define the basis function approximation as:
\begin{equation}
\hat{V}(x,\theta) = \theta^T\phi_\mathrm{c}(x)
\end{equation}
with $\phi_\mathrm{c}(x)$ a 3$^{\mathrm{rd}}$-order Fourier basis resulting in 16 learnable parameters $\theta$ in the domain $[q_{min}, q_{max}] \times [p_{min},p_{max}] = [-\pi,\pi] \times
[-8\pi J_p, 8\pi J_p]$.
Actor 1 ($\hat{P}_\mathrm{d}(q,\xi)$) is parameterized using a 3$^{rd}$-order Fourier basis in the range $[-\pi, \pi]$ resulting in 4 learnable parameters.
Actor 2 ($\hat{K}(x,\psi)$) is also parameterized using a 3$^{\mathrm{rd}}$-order Fourier basis for the full state space, in the same domain as the critic.
Exploration is done at every time step by randomly perturbing the action with a normally distributed zero-mean white noise with standard deviation $\sigma = 1$, i.e.:
\begin{equation}
\Delta u \sim \mathcal{N}(0,1)
\end{equation}
We incorporate saturation by defining the saturation function \eqref{eq:saturation} as:
\begin{equation}
\varsigma(u_k)=
\left\{
\begin{array}{ll}
u_k &\text{if}~\vert u_k \vert \le u_{\max} \\
\sgn(u_k)u_{\max} &\text{otherwise}
\end{array}
\right.
\label{eq:satfunction}
\end{equation}
Recall that the saturation must be taken into account in the policy gradients by applying \eqref{eq:polgradsat}-\eqref{eq:polgradsat1}.
The parameters were all initialized with zero vectors of appropriate dimensions, i.e. $(\theta_0,~\xi_0,~\psi_0) = 0$.
The algorithm was first run with the system simulated in Matlab for 200 trials
of three seconds each (with a near-optimal policy, the pendulum needs
approximately one second to swing up). Each trial begins in the initial position $x_0$. This simulation
was repeated 50 times to get an estimate of the average, minimum, maximum and
confidence regions for the learning curve. The simulation parameters are given
in Table~\ref{tab:simparam}.
\begin{table}[htbp]
\caption{Simulation parameters}
\label{tab:simparam}
\begin{center}
\begin{tabular}{lcll}
\hline
\textbf{Simulation parameters} &\textbf{Symbol} &\textbf{Value} &\textbf{Units} \\
\hline
Number of trials &$-$ &$200$ &- \\
Trial duration	&$T_\mathrm{t}$		 &$3$ &s \\
Sample time		&$T_\mathrm{s} $ &$0.03$ &s \\
Decay rate		&$\gamma$		 &$0.97$ &- \\
Eligibility trace decay	&$\lambda$		 &$0.65$ &- \\
Exploration variance	&$\sigma^2$ 		&$1$ &- \\
Max control input &$u_{\max}$ &$3$		 &$V$ \\
Learning rate of critic	&$\alpha_\mathrm{c}$ &$0.05$		 &- \\
Learning rate of $\hat{P}_\mathrm{d}(q,\xi)$	 &$\alpha_{\mathrm{a}_\mathrm{b},\xi} $ &$1\times 10^{-10}$		 &- \\
Learning rate of $\hat{K}(x,\psi)$	&$\alpha_{\mathrm{a}_\mathrm{b},\psi}$ &$0.2$		 &- \\
\hline
\end{tabular}
\end{center}
\end{table}
   \begin{figure}[t!]
      \centering
      \includegraphics[width=0.49\textwidth]{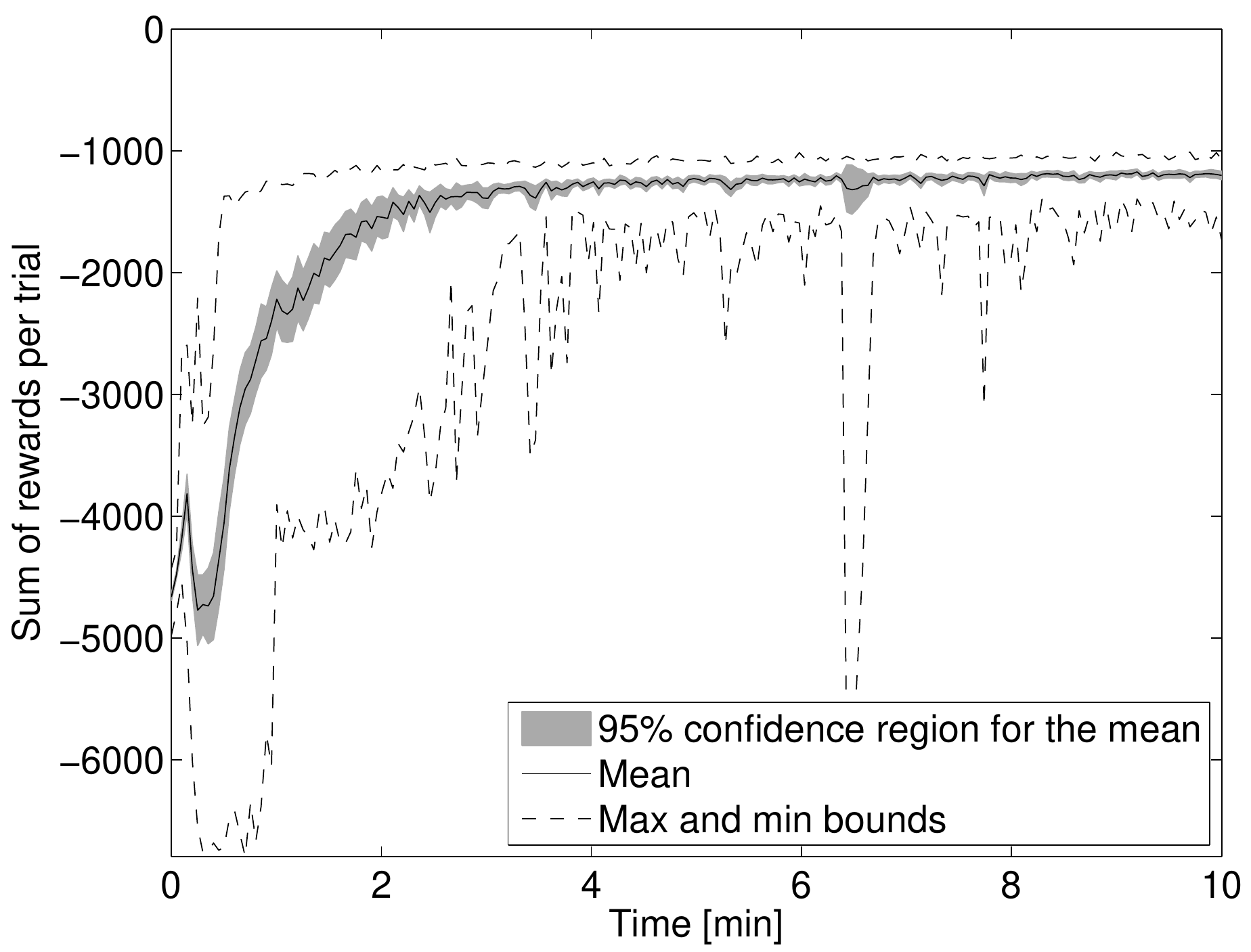}
      \caption{Results for the EBAC method for 50 learning simulations.}
      \label{fig:50sim}
   \end{figure}
   \begin{figure}[t!]
      \centering
      \subfloat[Simulated response]{\label{fig:sim}\includegraphics[width=0.49\columnwidth]{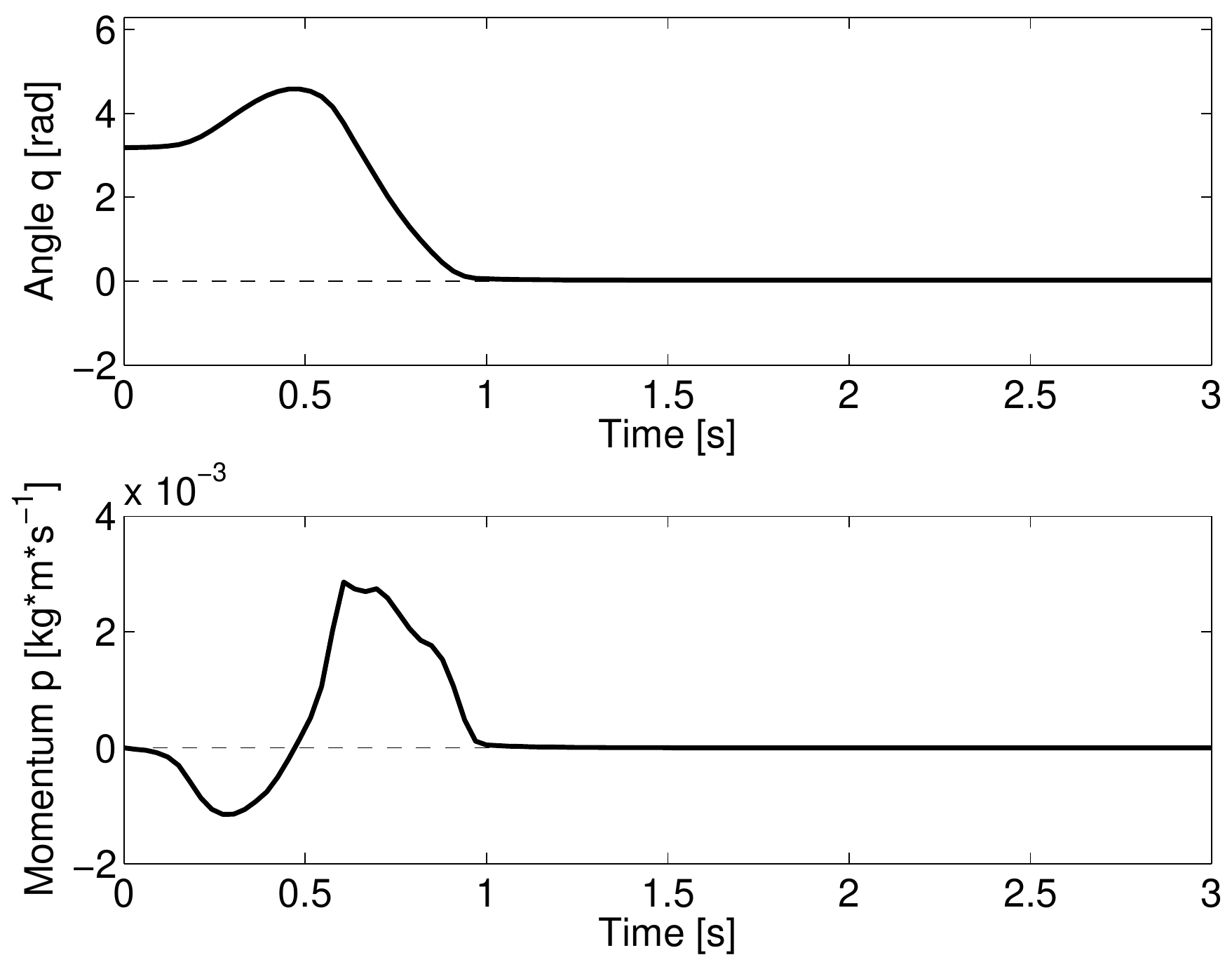}}
      \subfloat[Desired Hamiltonian]{\label{fig:HD}\includegraphics[width=0.49\columnwidth]{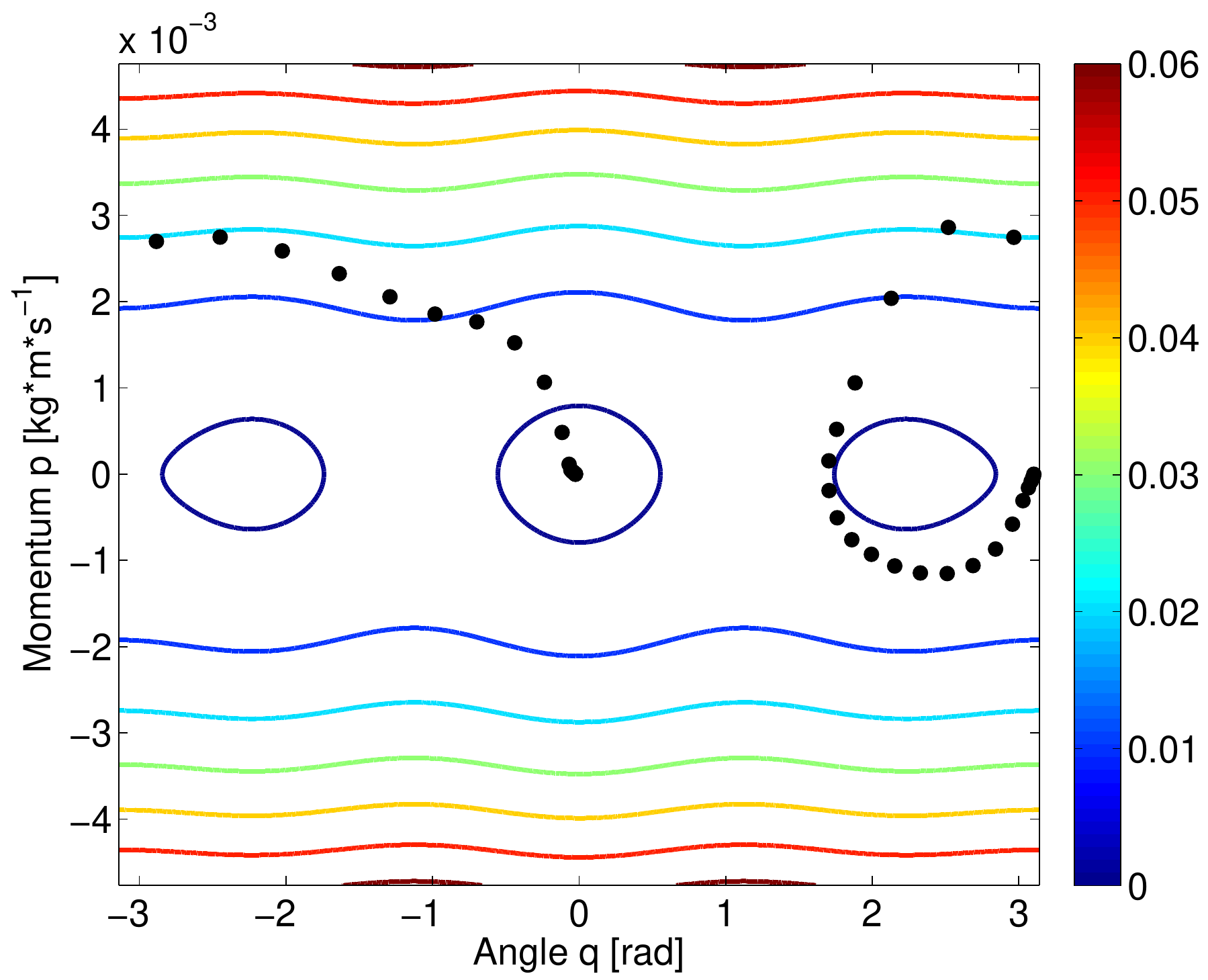}}
      \caption[Desired Hamiltonian and system response for a typical policy]{Simulation results for the angle
                $q$ (a, top), momentum $p$ (a, bottom) and the desired closed-loop Hamiltonian $H_\mathrm{d}(x,\xi,\psi)$ (b)
                including the simulated trajectory (black dots) using the policy learned.}
      \label{fig:HDandsim}
   \end{figure}
Fig.~\ref{fig:50sim} shows the average learning curve obtained after 50
simulations. The algorithm shows good convergence and on average needs about 2
minutes (40 trials) to reach a near-optimal policy. The initial drop in
performance is caused by the zero-initialization of the value function
(critic), which is too optimistic compared to the true value function.
Therefore, the controller explores a large part of the state space and receives
a lot of negative rewards before it learns the true value of the states. A
simulation using the policy learned in a typical experiment is given in
Fig.~\ref{fig:sim}. As can be seen, the pendulum swings back once to build up
momentum to eventually get to the equilibrium. The desired Hamiltonian
$\hat{H}_\mathrm{d}(x,\xi)$ \eqref{eq:deshampar2}, acquired through learning,
is given in Fig.~\ref{fig:HD}. There are three minima, of which one corresponds
to the desired equilibrium. The other two equilibria are undesirable wells that
come from the shaped potential energy $\hat{P}_\mathrm{d}(q,\xi)$
(Fig.~\ref{fig:Vd}). These minima are the result of the algorithm trying to swing up the pendulum in a single swing, which is not possible due to the saturation. Hence, a swing-up strategy is necessary to avoid staying in these wells. The number of these undesirable wells is a function of the control saturation and of the number of basis functions chosen to approximate $\hat{P}_\mathrm{d}(q,\xi)$. The learned damping $\hat{K}(x,\psi)$ (Fig.~\ref{fig:Kd})
is positive (white) towards the equilibrium thus extracting energy from the
system, while it is negative (gray) in the region of the initial state. The
latter corresponds to pumping energy into the system, which is necessary to
build up momentum for the swing-up and to escape the undesirable wells of
$\hat{P}_\mathrm{d}(q,\xi)$ (see discussion of expression (\ref{eq:gab1})). A disadvantage is that control law
\eqref{eq:uparam2-2}, with the suggested basis functions, is always zero for
the set $\Omega = \{x~\vert~ x= (0+j\pi, 0 ),~j=1,2,\dots\}$ which implies that
it is zero not only at the desired equilibrium, but also at the initial state $x_0$. During
learning this is not a problem because there is constant exploration, but after
learning the system should not be initialized in exactly $x_0$ otherwise it will stay
in this set. It can be overcome by initializing with a small perturbation
$\epsilon$ around $x_0$. In real-life systems it will also be less a problem
because there is generally noise present on the sensors.
   \begin{figure}[htbp]
      \centering
      \subfloat[$\hat{P}_\mathrm{d}(q,\xi)$]{\label{fig:Vd} \includegraphics[width=0.49\columnwidth]{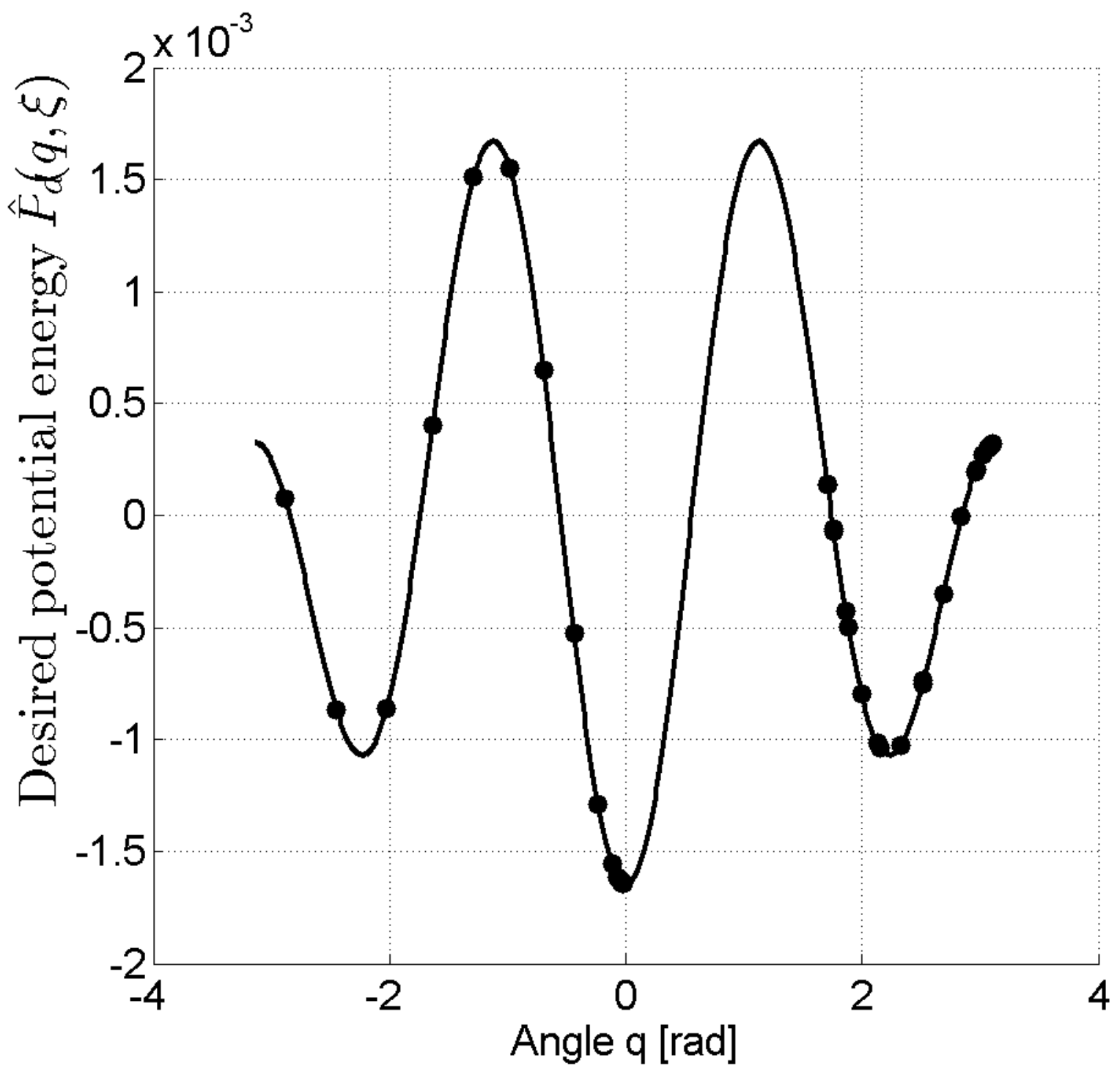}}
      \subfloat[$\sgn{\left(\hat{K}(x,\psi))\right)}$]{\label{fig:Kd} \includegraphics[width=0.49\columnwidth]{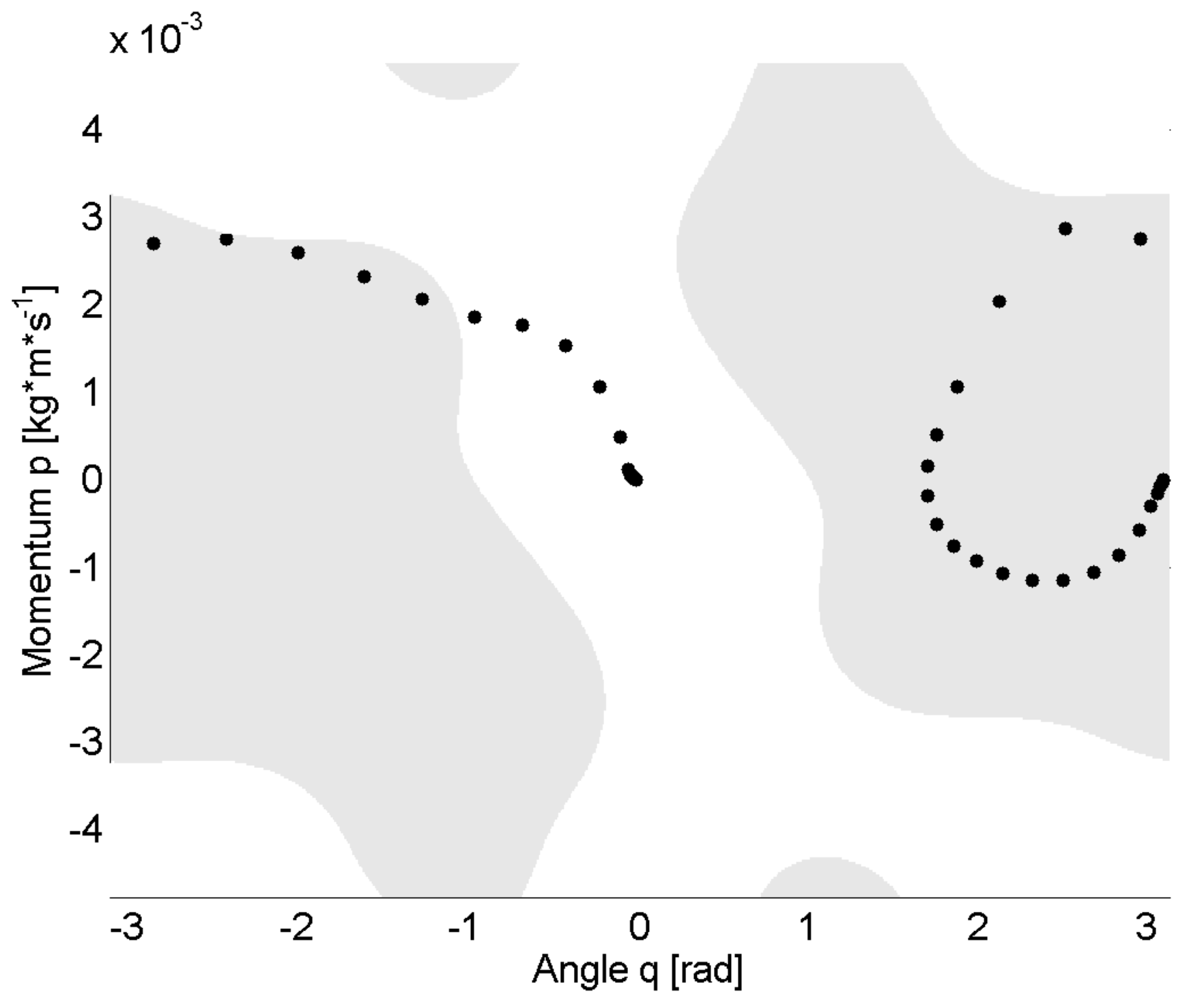}}
      \caption[Desired potential energy and desired damping]{Desired potential energy (a) and desired damping (b) (gray: negative; white: positive) for a typical learning experiment. The black dots indicate the value of the respective quantity for the simulation of Fig.~\ref{fig:sim}.}
      \label{fig:vdrd}
   \end{figure}

\subsection{Stability of the Learned Controller} \label{subsec:stab}

\begin{figure}[t!]
      \centering
      \subfloat[$\sgn{\left(\dot{\hat{H}}_\mathrm{d}(x,\psi)\right)}$]{\label{fig:Hddot} \includegraphics[width=0.49\columnwidth]{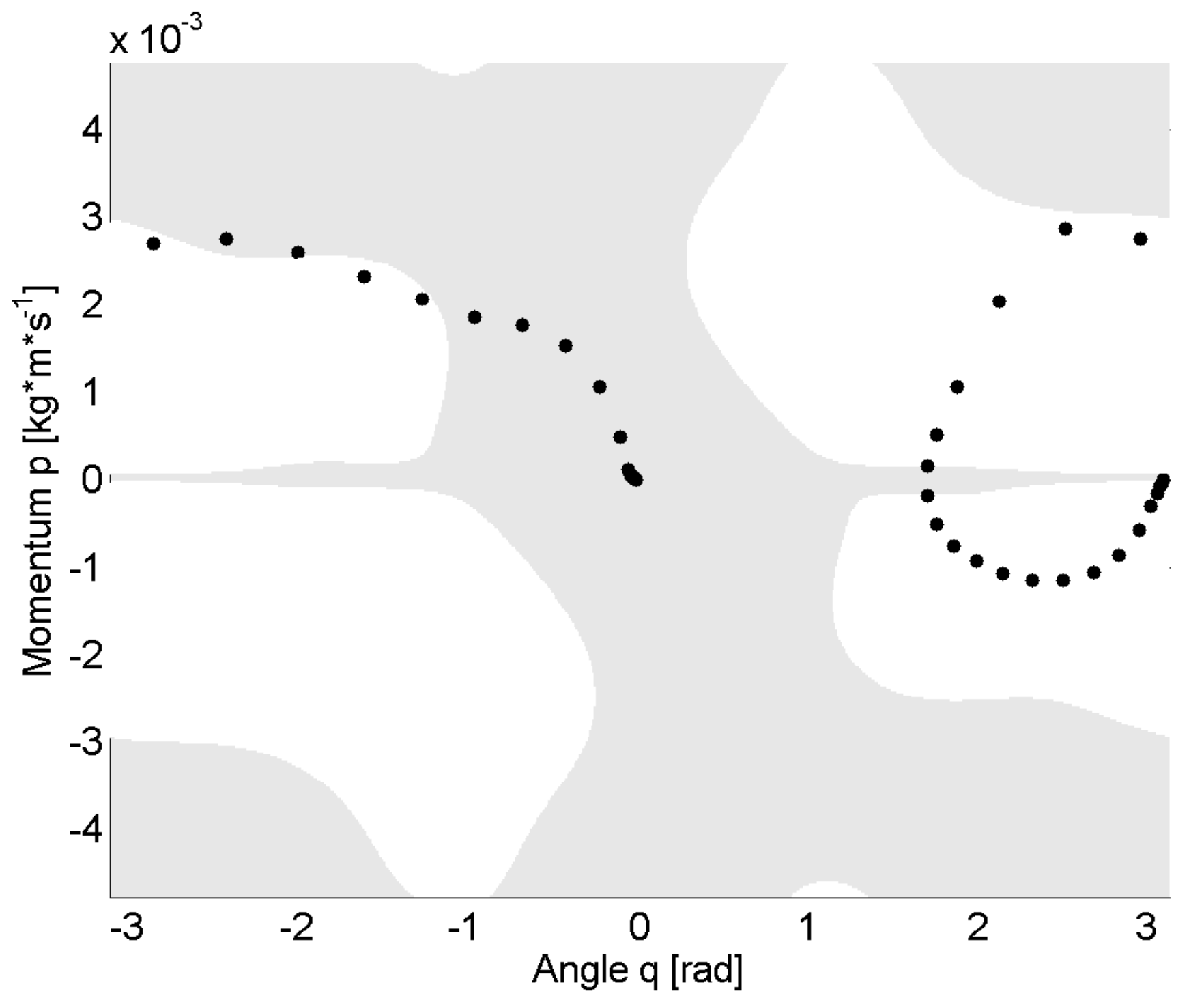}}
      \subfloat[$\dot{\hat{H}}_{\mathrm{d},\mathrm{diff}}(x,\psi)$]{\label{fig:satHddot} \includegraphics[width=0.49\columnwidth]{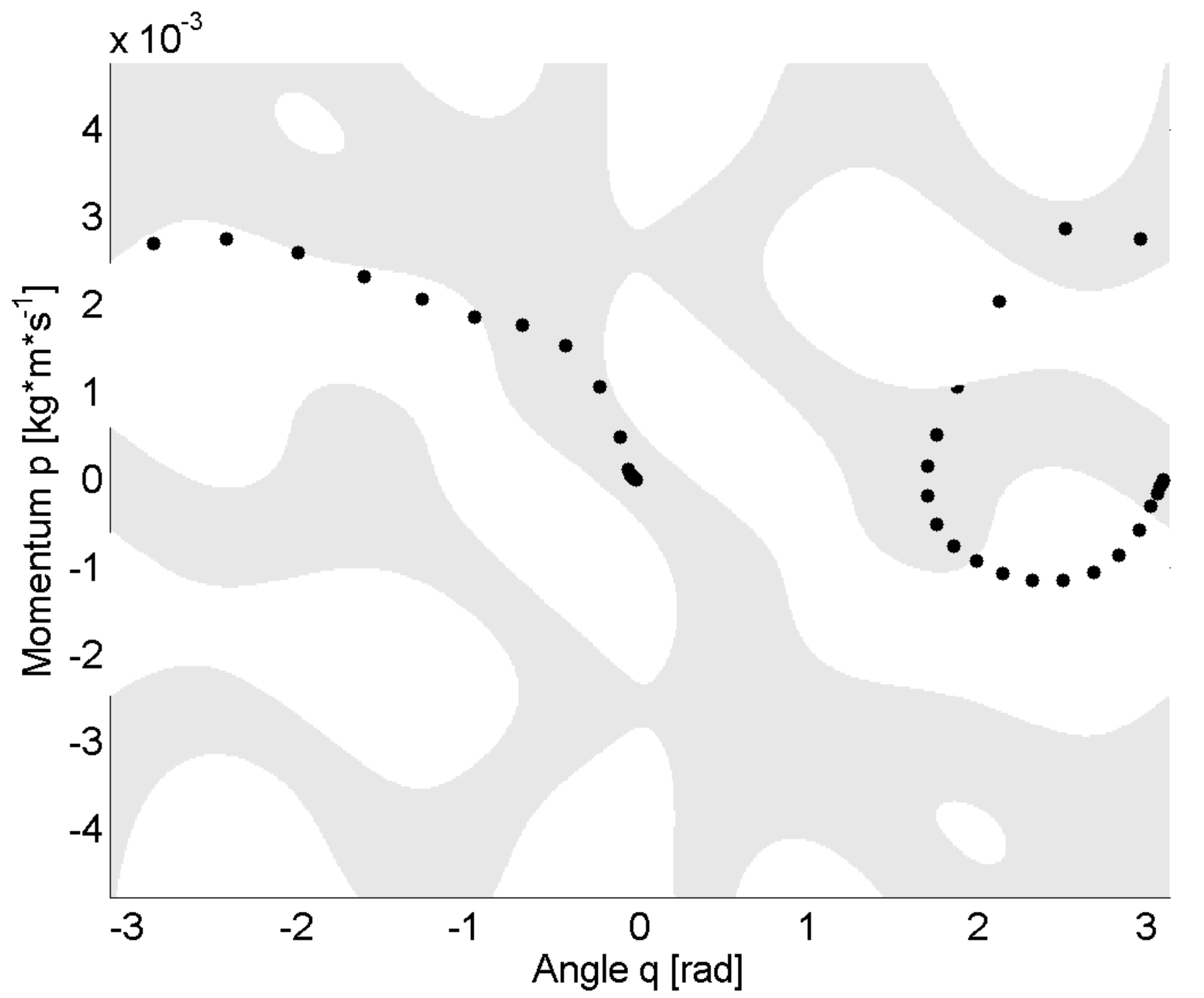}}
      \caption[Signum of $\dot{\hat{H}}_\mathrm{d}(x,\psi)$ indicating where $\dot{\hat{H}}_\mathrm{d}(x,\psi)=\dot{\hat{H}}_{\mathrm{d,sat}}(x,\psi)$]{Signum of $\dot{\hat{H}}_\mathrm{d}(x,\psi)$
      (a) indicating positive (white) and negative (gray) regions and (b) $\dot{\hat{H}}_\mathrm{d,diff}(x,\psi)= \sgn\left(\dot{\hat{H}}_\mathrm{d}(x,\psi)-\dot{\hat{H}}_{\mathrm{d},\sat}(x,\psi)\right)$ indicating regions where $\dot{\hat{H}}_\mathrm{d}(x,\psi)=\dot{\hat{H}}_{\mathrm{d},\sat}(x,\psi)$ (gray) and $\dot{\hat{H}}_\mathrm{d}(x,\psi)\neq\dot{\hat{H}}_{\mathrm{d},\sat}(x,\psi)$ (white).
      Black dots indicate the simulated trajectory.}
      \label{fig:hdsathd}
   \end{figure}

Since control saturation is present, the target dynamics do not satisfy
\eqref{eq:tdyn}. Hence, to conclude local stability of $x^*$ based on
\eqref{eq:dotHD}, we calculate $\dot{\hat{H}}_\mathrm{d}(x,\xi)$ for the
unsaturated case (Fig.~\ref{fig:Hddot})\footnote{Fig.~\ref{fig:Hddot} is
sign-opposite to Fig.~\ref{fig:Kd}, which is logical, because the negative
(positive) regions of $\hat{K}(x,\psi)$ correspond to negative (positive)
damping which corresponds to a positive (negative) value of
$\dot{\hat{H}}_\mathrm{d}(x,\xi)$ based on \eqref{eq:dotHD}.} and the saturated
case ($\dot{\hat{H}}_{\mathrm{d},\sat}(x,\xi)$) and compute the sign of the
difference (Fig.~\ref{fig:satHddot}). By looking at Fig.~\ref{fig:satHddot}, it
appears that $\exists \delta \subset \mathbb{R}^n:
\vert x-x^* \vert < \delta$ such that $\dot{\hat{H}}_{\mathrm{d},\sat}(x,\xi)  =
\dot{\hat{H}}_\mathrm{d}(x,\xi)$. It can be seen from Fig.~\ref{fig:satHddot}
that such a $\delta$ exists, i.e., a small gray region around the equilibrium
$x^*$ exists. Hence, we can use $\dot{\hat{H}}_\mathrm{d}(x,\xi)$ around $x^*$ and assess stability using \eqref{eq:dotHD}.
From Fig.~\ref{fig:HD} it follows that
$\hat{H}_\mathrm{d}(x,\xi) > 0$ for all states in Fig.~\ref{fig:HD}. From
Fig.~\ref{fig:Vd} we infer that locally, 
\begin{align}
\arg \min \hat{P}_\mathrm{d}(q,\xi) &=x^*;~~
\dot{\hat{P}}_\mathrm{d}(x^*,\xi) = 0;~~ 
\ddot{\hat{P}}_\mathrm{d}(x^*,\xi)  > 0 
\end{align}
the latter two of which naturally
result from the basis function definition. Furthermore, from Fig.~\ref{fig:Kd}
it can be seen that around $x^*$, $\hat{K}(x,\psi) > 0$. Hence, in a region
$\delta$ around $x^*$,
\begin{align}
\hat{H}_\mathrm{d}(x,\xi) 	      &> 0;~~
\dot{\hat{H}}_\mathrm{d}(x,\xi) \le 0;~~
\dot{\hat{H}}_\mathrm{d}(x^*,\xi) =  0
\end{align}
which implies local asymptotic stability of $x^*$. Extensive simulations show
that similar behaviour is always achieved.

\subsection{Real-time Experiments}

\begin{figure}[htbp]
      \centering
      \includegraphics[width=0.49\textwidth]{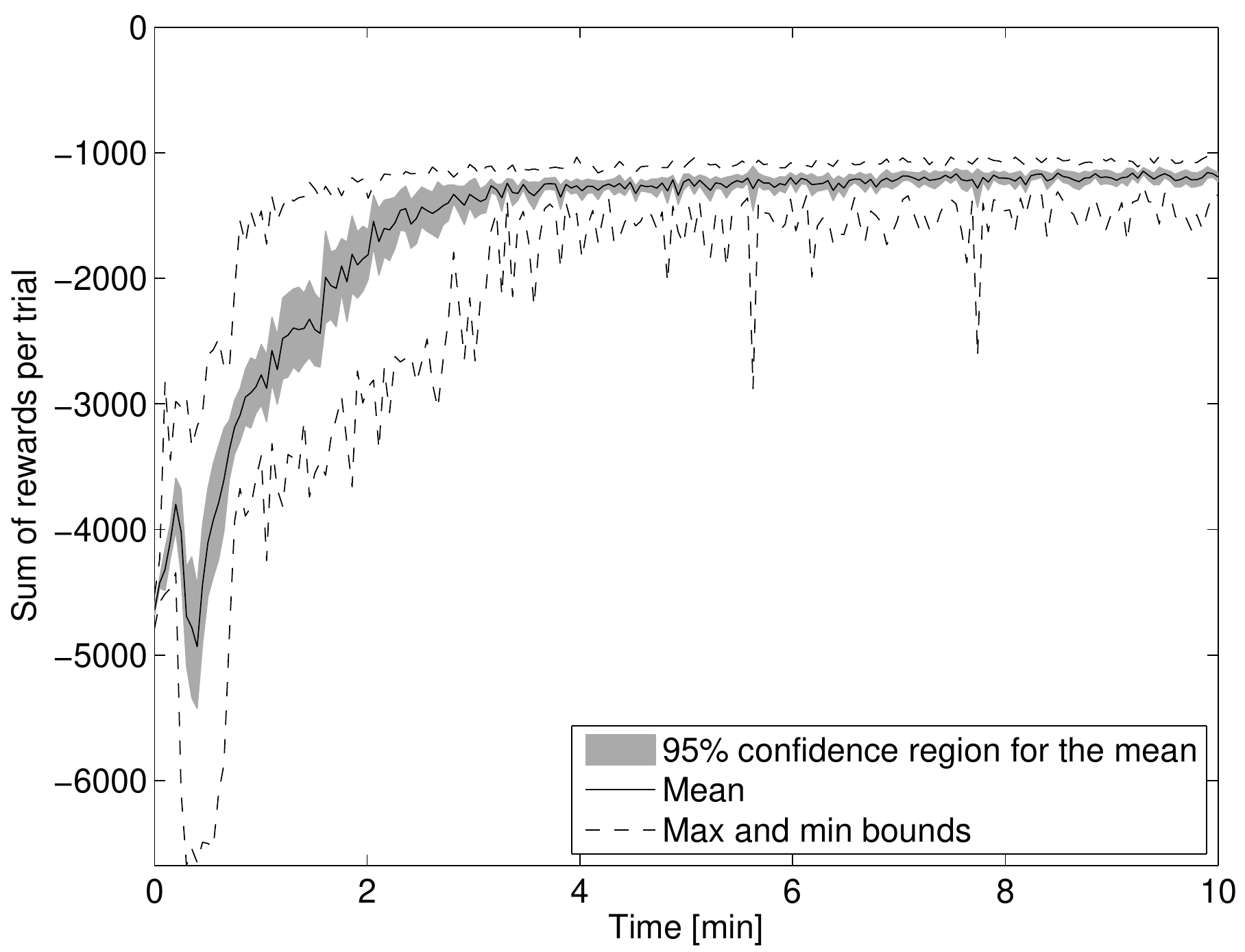}
      \caption{Results for the EBAC method for 20 learning experiments with the real physical system.}
      \label{fig:20exp}
   \end{figure}

Using the physical setup shown in Fig.~\ref{fig:pendulum}, 20 learning experiments were
run using identical settings as in the simulations. The result is given in
Fig.~\ref{fig:20exp}. The algorithm shows slightly slower convergence - about 3
minutes of learning (60 trials) to reach a near-optimal policy instead of 40 -
and a less consistent average when compared to Fig.~\ref{fig:50sim}. This can
be attributed to a combination of model mismatch and the symmetrical basis
functions (through which it is not possible to incorporate non-symmetrical
friction that is present in the real system). Overall though, the performance can be considered good when
compared to the simulation results. Also, the same performance dip is present
which can again be attributed to the optimistic value function initialization.

\section{Conclusions} \label{sec:CO}

In this paper, we have presented a method to systematically parameterize EB-PBC
control laws that is robust to extra nonlinearities such as input control saturation. The parameters are then found by making use of actor-critic
reinforcement learning. In this way, we are able to learn a closed-loop energy
landscape for PH systems. The advantages are
that optimal controllers can be generated using
energy-based control techniques, there is no need to specify a global system Hamiltonian, and the solutions acquired by means of
reinforcement learning can be interpreted in terms of energy shaping and
damping injection, which makes it possible to numerically assess stability
using passivity theory. By making use of the model knowledge the actor-critic
method is able to quickly learn near-optimal policies. A drawback is that for
multiple input systems, generating many actor updates for the desired damping
matrix can be computationally expensive. We have found that the proposed Energy Balancing Actor Critic algorithm performs very well in a physical mechanical setup. Due to the intrinsic energy boundedness of the learned desired Hamiltonian, we have observed that the system never gets unstable during learning. We are currently active on the extension of the algorithms presented to an IDA-PBC setting, such that more classes of systems can be addressed and more freedom is given in shaping the desired Hamiltonian (e.g. Kinetic energy shaping).

\ifCLASSOPTIONcaptionsoff
  \newpage
\fi

\bibliographystyle{IEEEtran}
\bibliography{bib_thesis}

\end{document}